\documentclass[epj]{svjour}
\usepackage{graphics}
\usepackage{amsmath}
\usepackage{color,widetext}
\newcommand{\mean}[1]{\ensuremath{\langle#1\rangle}}

\newcommand{\ket}[1]{\ensuremath{\lvert#1\rangle}}
\DeclareMathOperator{\atan}{atan}
\begin{document}
\title{Dissipative dynamics of a quantum two-state system in presence of nonequilibrium quantum noise}
\titlerunning{Nonequilibrium quantum two-state system}
\author{Niklas Mann, Jochen Br\"uggemann
 \and Michael Thorwart
}                    
\institute{I. Institut f\"ur Theoretische Physik, Universit\"at Hamburg,
Jungiusstra{\ss}e 9, 20355 Hamburg, Germany}
\date{Received: date / Revised version: date}
\abstract{We analyze the real-time dynamics of a quantum two-state system in the presence of nonequilibrium quantum fluctuations. The latter are generated by a coupling of the two-state system to a single electronic level of a quantum dot which carries a nonequilibrium tunneling current. We restrict to the sequential tunneling regime and calculate the dynamics of the two-state system, of the dot population, and of the nonequilibrium charge current on the basis of a diagrammatic perturbative method valid for a weak tunneling coupling. We find a nontrivial dependence of the relaxation and dephasing rates of the two-state system due to the nonequilibrium fluctuations which is directly linked to the structure of the unperturbed central system. In addition, a Heisenberg-Langevin-equation of motion allows us to calculate the correlation function of the nonequilibrium fluctuations. By this, we obtain a generalized nonequilibrium fluctuation relation which includes the equilibrium fluctuation-dissipation theorem. A straightforward extension to the case with a time-periodic ac voltage is shown.  
} 
\maketitle
%
\section{Introduction}\label{sec:1}
The loss of quantum coherence of a pure superposition state of a quantum system has been in the focus of research since the early days of quantum physics
\cite{Bogolyubov45,Weiss,Zwerger,Grabert,Kubo}. Quantum decoherence, and with it, quantum relaxation, are at the heart of the fundamental question how quantum mechanics is reconciled with the appearance of a classical world \cite{ZehBook}. A modeling of these processes needs to include quantum statistical fluctuations in the modeling of the dynamics of a quantum system \cite{Bogolyubov45,Kubo}. A simple way of such a model is to consider an infinite set of uncoupled quantum harmonic oscillators held at thermal equilibrium (bath or environment) and coupled to the central system of interest. Such an equilibrium bath generates fluctuations for the system leading to quantum dephasing and dissipation. A minimal model of a central system to study these questions is a quantum two-state model in which the two states can dephase and perform dissipative transitions due to the coupling to the bath \cite{Weiss,Zwerger}. This (equilibrium) spin-boson model forms a cornerstone of quantum statistical physics and shows fascinating non-trivial behaviour, such as a quantum phase transition to a localized phase in which tunneling between the two states is suppressed by strong dissipation. Likewise, it has served as a key model for the development of many fundamental numerical and analytical methods to describe quantum dissipative dynamics. 

Among many of these methods, real-time path integral approaches \cite{Weiss,Zwerger} allow to obtain analytical results for the spin-boson model in certain limiting cases. In the limit of a strong coupling to the equilibrium fluctuations, the non-interacting blip approximation (NIBA) allows to treat the transitions between the two states perturbatively and to reveal a mainly incoherent decay of the population and the coherence. In contrast, a weak coupling of the two states to the fluctuations allows a perturbative expansion in the damping constant \cite{Weiss} such that the dephasing and the relaxation rates follow in the form of a simple analytic expression. There, the coherent transitons between the two states are treated to all orders. Again, the quantum dissipative dynamics is due to equilibrium (quantum) fluctuations of the bath.

An important physical principle valid at thermal equilibrium is the (quantum) fluctuation-dissipation theorem \cite{Callen,Kubo} which connects the action of (quantum) fluctuations with the inevitable consequence of the appearance of (quantum) dissipation. It is well known that this elegant principle does not hold for nonequilibrium fluctuations and efforts are made to find a general principle for those as well \cite{NJPFocusIssue15}.

The action of nonequilibrium quantum fluctuations to a coherent quantum two-state system is more complex and no closed exact expressions for the relaxation and the dephasing rates exist up to present. A simple model to study the impact of nonequilibrium quantum statistical fluctuations is a quantum two-state (or spin-1/2) system which is coupled to the energy level of a single spinless non-interacting electron on a quantum dot. The electron can tunnel from or into reservoirs of non-interacting electrons held at a common constant temperature. A straightforward way to achieve nonequilibrium conditions is to use at least two reservoirs (a 'left' and a 'right' one), between which a constant electric voltage is applied. The finite difference of the electrochemical potential gives rise to a charge transport between the two leads via the quantum dot which induces nonequilibrium quantum fluctuations to the quantum two-state system coupled to the dot. Although the two electron reservoirs are fermionic in nature, they are free non-interacting particles, such that the situation resembles the equilibrium spin-boson model to some extent. Despite the statistical distribution functions are of course different, this model has been called the nonequilibrium spin-boson model \cite{Mitra05,Segal2007}. 

This nonequilibrium quantum two-state model has been addressed in Ref.\ \cite{Mitra05} on the basis of a NIBA-like approximation. The central system Hamiltonian is $H_S=BS_z+\Delta S_x+JS_z d^\dagger d$ with $S_i=\sigma_i/2$, with Pauli matrices $\sigma_{i=x,y,z}$ and $d$ being the fermion operator of the dot electron level which is tunnel-coupled to non-interacting leads. In the case of vanishing off-diagonal coupling in the quantum two-state system, $\Delta \to 0$, the full problem becomes exactly solvable. Hence, it is reasonable to study the effect of $\Delta$ perturbatively. Under the additional assumption of not too strong system-bath coupling $J$, an expansion of the noise correlators up to third order in $J$ can be used. Then, the frequency-dependent spin-relaxation rate constants, the frequency-dependent fluctuation-dissipation ratio and an effective frequency-dependent nonequilibrium 'temperature' could be determined \cite{Mitra05}. 

A further extended analysis of the regime of small $\Delta$ characterized by 
 a golden rule rate for the transitions between the two energy eigenstates proportional to $\Delta^2$ has been carried out in Ref.\ \cite{Segal2007}. The intermediate time domain has been carefully addressed for the full parameter regime of weak to strong system-bath coupling $J$ at zero (or very low) temperature. A Marcus-like nonequilibrium quantum relaxation rate has been derived. Interestingly, a simple mapping between the equilibrium temperature and bias voltage has been shown not to exist. Different decay characteristics in different time regimes involve algebraic as well as exponential decays of the correlation function. In addition, going beyond the lowest order in  
 $\Delta$, it has been shown that a Coulomb gas behavior in terms of a power series in 
 $\Delta$ is suggested to be valid, since the first few orders up to the contribution $\sim \Delta^6$ agree with the numerically calculated exact result. Yet, a complete analysis to all orders in $\Delta$, but for linear order in $J$ at finite temperature remained open. 
 
A variant of this model, where a quantum two-state system is coupled directly to two bosonic baths held at different temperatures (without involving fermions), has been set up to study heat exchange \cite{Nicolin2011}. On the basis of a NIBA-like treatment, the full counting statistics via the cumulant generating function has been calculated and  fluctuation relations have been discussed. Yet, the exchange of energy or heat rises fundamental questions of the notion of heat under nonequilibrium conditions \cite{Galperin}. 

In this work, we address the dynamics of the two-state system under the influence of nonequilibrium quantum fluctuations provided by an electronic charge flowing between two non-interacting leads in the regime of weak system-bath (tunneling) coupling. We provide an analysis in terms of lowest-order tunneling processes between quantum dot and leads. The coupling between the electronic level on the dot and the quantum two-state system can be arbitrarily large. The weak tunneling coupling to the fluctuations generates only a small broadening and a small energy shift of the system states such that the unperturbed energy spectrum is a good starting point. We employ the well established diagrammatic perturbation method \cite{Schoeller96,PhysRevB.54.16820} formulated on the Keldysh contour to determine the real-time dynamics of the quantum two-state system under the action of a nonequilibrium charge current. The population of the central dot and the flowing charge current follow as well. This allows us to extract the relaxation and dephasing rate for the central two-state system and to explain their dependence on the various parameters in terms of the unperturbed energy spectrum. In addition, we use the Heisenberg-Langevin equations of motion to determine the autocorrelation function of the nonequilibrium fluctuations. For a static (dc) bias voltage between the two leads, we find a generalized nonequilibrium fluctuation relation which extends the well-known equilibrium fluctuation-dissipation theorem. The Fourier transform of the autocorrelation function includes Ohmic contributions as well as nontrivial Lorentzian terms. Due to its simple structure, the nonequilibrium fluctuation relation can be generalized to time-periodic (ac) transport voltages and leads us to a Floquet-fluctuation relation in terms of higher harmonics of the correlation function of the nonequilibrium noise. 

\section{Model}
\label{sec:2}
We model the quantum mechanical two-state system which is coupled to
nonequilibrium quantum fluctuations provided by a flowing electron current
through a quantum dot. To be specific, we couple the two quantum states of
interest $\ket{\uparrow}$ and $\ket{\downarrow}$ to the time-dependent
population of electrons on the dot. The corresponding destruction and creation
operators of a single electron are denoted as $d$ and $d^\dagger$. 
Formally, this is described  by the system Hamiltonian
\begin{equation}
 H_S=BS_z+\Delta S_x+JS_zd^\dagger d
\label{eq:Hsys}
\end{equation}
in spin representation $S_i=\sigma_i/2$, with Pauli matrices $\sigma_{i=x,y,z}$.
The energy difference between the two states is given by $B$, and we allow a
transition with a coupling constant $\Delta$. The effective spin$-1/2$ is
coupled to the electronic occupation number operator $n=d^\dagger d$ of the
dot with a coupling strength $J$. The system Hilbert space is spanned by the
product basis $\{\lvert\sigma n\rangle\}$, where $\sigma=\uparrow, \downarrow$
and $n=0,1$ is the dot occupation number. 
The spectral decomposition reads $ H_S\lvert{\pm
n}\rangle=\epsilon_n^\pm\lvert{\pm n}\rangle$, where
$\epsilon_n^\pm=\pm\sqrt{\Delta^2+(B+nJ)^2}/2$ and $+(-)$ corresponds to an
(anti-)symmetric linear superposition of the two spin states.\\

The nonequilibrium fluctuations are generated by a flowing electron current
through the dot, such that the total system-bath
Hamiltonian has the form $ H= H_{S}+ H_{B}+ H_{T}$. The nonequilibrium
bath to which the system couples is composed of two non-interacting
reservoirs of spinless electrons with energies 
$\varepsilon_{k\alpha}$ described by the Hamiltonian 
\begin{equation}
 H_B=\sum_{k\alpha}\left(\varepsilon_{k\alpha} - \mu_\alpha \right) 
c_{k\alpha}^\dagger c_{k\alpha} \,
,
\end{equation}
 with reservoir index $\alpha=L/R$.  As usual, we assume
the reservoirs to be large enough so that the interaction with system has
almost no backaction on the fermions in the leads. The two baths are assumed to 
be thermalized at all times at the same temperature $T$, but at two different
chemical potentials $\mu_{L/R}$. We assume a
symmetric voltage bias between both reservoirs, i.e., $\mu_L-\mu_R=eV$.

The interaction between the dot states and the reservoirs is described via
a tunneling coupling Hamiltonian 
\begin{equation}
 H_{T}=\sum_{k\alpha} t_{k\alpha}d^\dagger c_{k\alpha}+h.c.,
\end{equation}
 with the tunneling amplitude $t_{k\alpha}$. The electronic operators $d$ and
$c_{k\alpha}$ obey the standard algebra of fermonic operators. Throughout this
work, we use natural units and set $\hbar=1,\,k_B=1$.
%
%
\section{Spin Dynamics and Electron Current}\label{sec:3}
We study the nonequilibrium quantum dynamics of the central system numerically
by using the well established diagrammatic perturbation method 
\cite{Schoeller96,PhysRevB.54.16820}. The time evolution of the expectation
value of an arbitrary operator $A$ is thereby written in terms of a
time-ordered integration along the Keldysh contour $K$
\begin{align}\begin{split}
\mean{ A(t)}=&\text{tr}{\biggl[\rho_0\mathcal{T}_K\exp{
\left(-i\int_Kd\tau H_{T}(\tau)\right)} A(t)\biggr]},\label{eq:meq}
\end{split}\end{align}
with the time-ordering operator $\mathcal{T}_K$ acting on the Keldysh contour. 
Any higher correlation function correspondingly follows. The time dependence 
of all operators on the right-hand side is meant as the free time evolution.
Assuming that the initial density operator $\rho_0$ factorizes into system and
bath part, one can
use equation \eqref{eq:meq} to find a systematic expansion organized in orders
of $H_{T}$. By choosing $A(t)$ to be the projection operator
$(\lvert i\rangle\langle j\rvert)(t)$, where $\{\lvert i\rangle\}$ is a set of
states that spans the system Hilbert space, one can calculate the time-evolution
of the system (or reduced) density matrix $\bf P$, i.e.,  the mean value of the
projection operator. The equation of motion for the reduced density operator is
obtained by inserting the projection operator and differentiating with respect
to time. One can rewrite the result as
\begin{equation}
 \partial_t{\bf P}={\bf WP}.\label{eq:meqp}
\end{equation}
The time-evolution operator $\bf W$ follows from expanding the exponential in
equation \eqref{eq:meq}. In a similar fashion we can find an equation that
allows us to
determine the tunnel current $I=e\partial_t(N_L-N_R)$, where $N_\alpha$ is the
particle number operator of lead $\alpha$. The stationary current follows
\cite{Schoeller96,PhysRevB.54.16820} as 
\begin{equation}
 I=\frac{e}{2}\mean{{\bf W}^I}_\text{st},\label{eq:curr}
\end{equation}
where the mean is taken with respect to the stationary solution of Eq.\ 
\eqref{eq:meqp}. Both $\bf W$ and ${\bf W}^I$ can be determined by using a
diagrammatic representation of the perturbation series. Explicit rules for
these diagrams are given 
\cite{PhysRevB.54.16820,PhysRevB.68.115105}.

When expanding the perturbation series, one has to estimate the mean value of
the current, and since the total number of electrons is conserved, only even
orders in $t_{k\alpha}$ contribute. Below, we use lowest order
perturbation in the tunnel amplitude, which is the second order in
$t_{k\alpha}$. Hence, it is useful to encode the tunneling strengths in the
parameter
$\Gamma_{\alpha}={2\pi}\sum_{k}|{t_{k\alpha}}|^2\delta(E_{F, \alpha}-\epsilon_{
k\alpha})$, where $E_{F, \alpha}$ is the Fermi energy of lead $\alpha$. 
We focus in this work on symmetric tunneling geometries and set 
$\Gamma_L=\Gamma_R=\Gamma$. Moreover, we use the standard wide-band
limit, which corresponds to the case where the amplitude is energy
independent around the Fermi energy. 

\subsection{Nonequilibrium Spin Relaxation}
First, we address the relaxation dynamics of the quantum two-level system
described by the spin $\bf S$. Expanding the density operator in terms of
the right eigenvectors of the non-Hermitian superoperator $\bf W$, we find the
solution to the Eq.\ \eqref{eq:meqp} as 
\begin{equation}
 {\bf P}(t)=\sum_{k=0}^{N^2-1} c_k{\bf
p}_ke^{-\Gamma_kt}e^{i\Omega_kt},\label{eq:denstev}
\end{equation}
where ${\bf p}_k$ is the $k$-th right eigenvector of $\bf W$ with eigenvalue
$\Lambda_k=-\Gamma_k+i\Omega_k$ and $N=4$ being the dimension of
the system Hilbert space. One eigenvalue $\Lambda_0$ is
equal to zero,
corresponding to the stationary solution. The eigenvalue with the smallest
non-zero real part yields the (asymptotic) relaxation rate $\Gamma_1$ of the
system. The coefficients $c_k$ have to be chosen such that
Eq.\ \eqref{eq:denstev} satisfies the initial condition ${\bf P}(0)=\sum_k c_k
{\bf p}_k$. 
%
%
\begin{figure}[t]\centering
\resizebox{0.49\textwidth}{!}{%
  \includegraphics{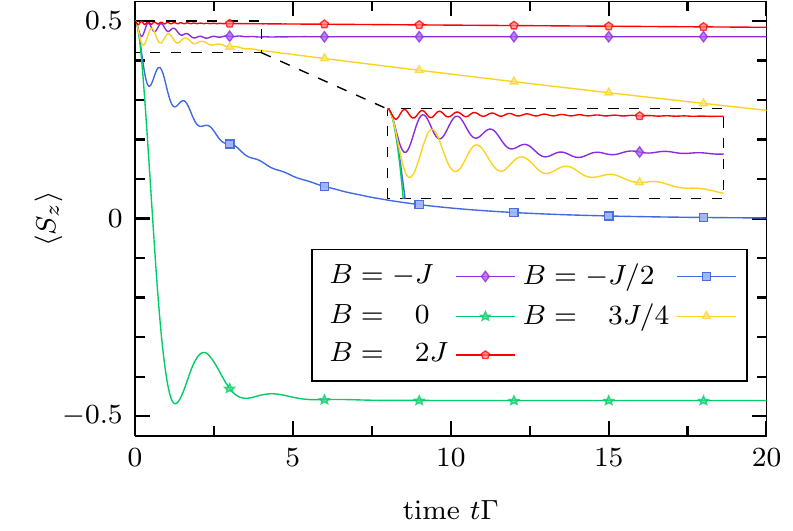}
}
\caption{Time-evolution of $\mean{S_z(t)}$ for the two
spin states $\ket{\uparrow}, \ket{\downarrow}$. The quantum two-level system 
is initially prepared in the state $\lvert\uparrow\rangle$ and the dot is
unoccupied. The parameters are $J=15\,\Gamma$, $\Delta=3\,\Gamma$,
$eV=10\,\Gamma$, and $\beta=0.7/\Gamma$. Moreover, we have chosen five
different values of $B$ according to $B=-\,J=-5\,\Delta=-15\,\Gamma$,
$B=-J/2=3\Delta/2=7.5\,\Gamma$, $B=3J/4=15\Delta/4=11.25\,\Gamma$ or
$B=2\,J=10\,\Delta=30\,\Gamma$. The inset shows a zoom to the short-time region
as indicated.}
\label{fig:tev}       
\end{figure}
%
%

\begin{figure}[t!]
\centering
\resizebox{0.49\textwidth}{!}{%
  \includegraphics{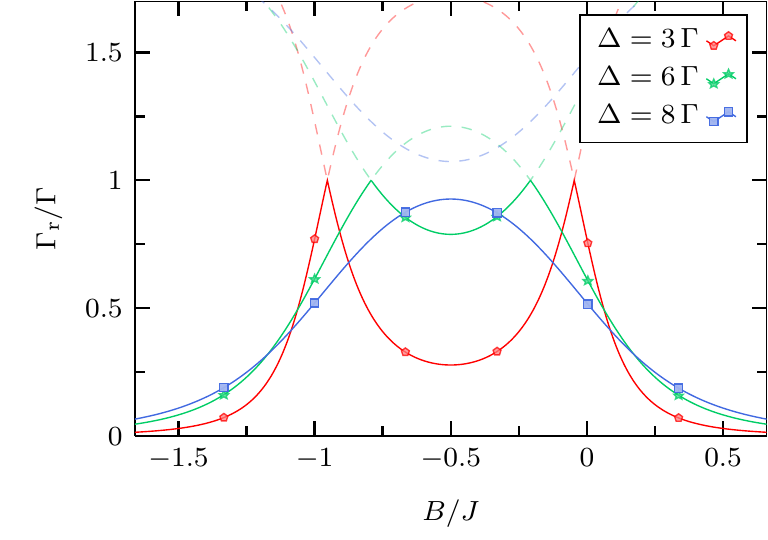}
}
\caption{Relaxation rate $\Gamma_r$ as a function of the energy splitting $B$
for different values of $\Delta$. The solid
lines show the smallest non-zero real eigenvalue of the Liouville
superoperator. Due to a degeneracy of two eigenvalues, an exact crossing of
the eigenvalues shapes the relaxation rate ragged at these points. The blurry
lines show the second smallest real eigenvalue. The remaining parameters are 
$J=15\,\Gamma$, $\Delta=3\,\Gamma$, $eV=10\,\Gamma$, and
$\beta=0.7/\Gamma$. }
\label{fig:rel2}       
\end{figure}

Fig.~\ref{fig:tev} shows the time-evolution of $\mean{S_z(t)}$ of the two
quantum states of the quasi-spin for different level splittings $B$ under the
influence of the nonequilibrium fluctuations provided by the electron current in
the quantum dot. In all five cases the spin relaxes to its asymptotic steady
state. The transient dynamical behavior shows decaying oscillations which are
charaterized by different oscillation frequencies and amplitudes. The
asymptotic approach to the steady state is monotonous and will be charaterized
by a single relaxation rate. The dynamical behavior is naturally determined by  
the energy spectrum of the system Hamiltonian, see below for a detailed
discussion. The relaxation is faster, when $\Delta$ is of the order of $B$ or
$B+J$. 
To determine the relaxation rate $\Gamma_1$ systematically, we calculate 
the smallest non-zero real eigenvalue according to Eq.~\eqref{eq:denstev}. The
result is shown in Fig.\ \ref{fig:rel2} for three different values of $\Delta$.
In the regime of large voltage ($eV=10 \Gamma$) addressed here, the relaxation
process is essentially independent of the temperature $T=1/\beta$ and
the voltage $eV$. Hence, the nonequilibrium relaxation rate $\Gamma_1$ is 
only a function of the system parameters $\Delta$, $B$, $J$, and the tunnel rate
$\Gamma$. For small $\Delta \sim \Gamma$, we find that the relaxation rate is
peaked at $B=0$ and $B=-J$ and shows a local minimum around $B=-J/2$. For larger
$\Delta$, the local maxima disappear and a global maximum at $B=-J/2$ arises.
The particular ragged behavior at the local maxima is due to a degeneracy of
two eigenvalues of the Liouville superoperator. In fact, each eigenvalue
depends monotonically on $B$, but due to an exact crossing at particular values
of $B$, the two eigenvalues exchange their role as minimal ones. 

\begin{figure}\centering
\resizebox{0.49\textwidth}{!}{%
  \includegraphics{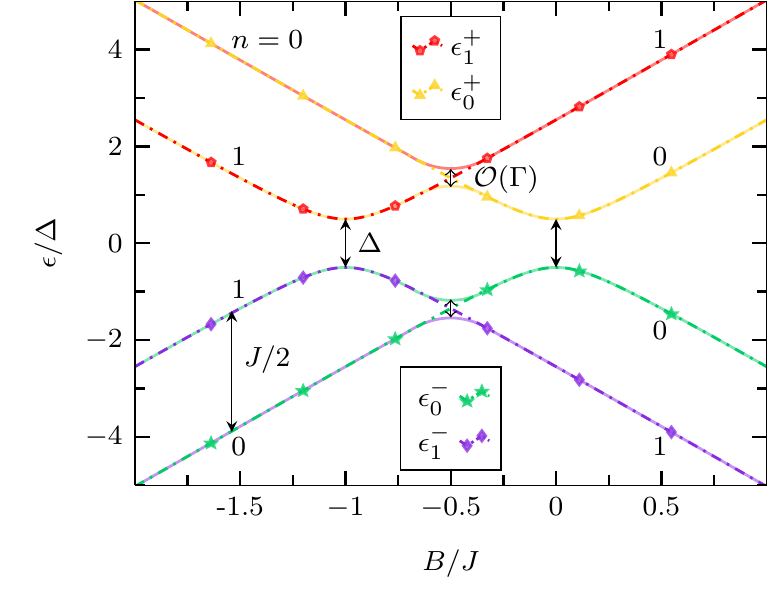}
}
\caption{Energy spectrum $\epsilon_n^\pm=\pm\sqrt{\Delta^2+(B+nJ)^2}/2$ of the
isolated quantum dot with electron occupation numbers $n=0,1$ for 
$J=5\Delta$.}
\label{fig:avcr}       
\end{figure}

To understand this behavior, it is necessary to consider the spectrum of the
system Hamiltonian. The energy levels as a function of $B$ are shown in Fig.\ 
\ref{fig:avcr} for $J=5\Delta$, which corresponds to the case $\Delta=3\
\Gamma$ in Fig.\ \ref{fig:rel2} (yellow line) and to the cases shown in 
Fig.\  \ref{fig:tev}. The two pairs of energy levels correspond to the two
possible electron occupation numbers $n=0,1$. At $B=-J/2$, two avoided level
crossings appear, where the associated energy gaps are determined by the
tunneling strength $\Gamma$. At these avoided crossings, two spin states with
different electron numbers mix, i.e., with $n=0$ and $n=1$.  The large energy
gap in between the two pairs is given by $J/2$. The two families of eigenstates
$n=0$ and $n=1$ are also separated in energy by $J/2$ far away from the avoided
crossings. Due to the coupling $\Delta$ of the two spin states, two more avoided
level crossings appear. The one at $B=0$ connects two states with electron
number $n=0$ on the dot. The one at $B=-J$ mixes the two spin states with
electron number $n=1$. The associated energy gaps are given by $\Delta$ at the
avoided crossing. 

%
%
%
%
%
%

Equipped with the energy spectrum, the behavior of the relaxation rate shown in
Fig.\ \ref{fig:rel2} can be understood. For $\Delta=3\Gamma$, the two local
ragged maxima occur due to the avoided level crossings at $B=0$ and
$B=-J$ between states with the same electron number, i.e., with $n=0$ at $B=0$
and $n=1$ at $B=-J$. Here, the dot occupation is stable and the fluctuations
are thus very efficient in relaxing the spin on the dot. In turn, at
$B=-J/2$, states with different electron numbers $n=0$ and $n=1$ on the dot are
mixed at the avoided crossings. This renders the fluctuations around an already
undetermined dot state inefficient and the corresponding relaxation rate
becomes minimal. \\
For growing $\Delta$, the gaps at $B=0$ and $B=-J$ become larger and the
avoided level crossings move towards each other. Eventually, the four avoided
level crossings are very close to each other around $B=-J/2$ such that they
cannot be resolved anymore when the reservoir parameters are kept fixed. Hence,
the two local maxima in the relaxation rate merge and form one broad peak
around $B=-J/2$, see Fig.\ \ref{fig:rel2}. 

%
\begin{figure}[t!]\centering
\resizebox{0.49\textwidth}{!}{%
  \includegraphics{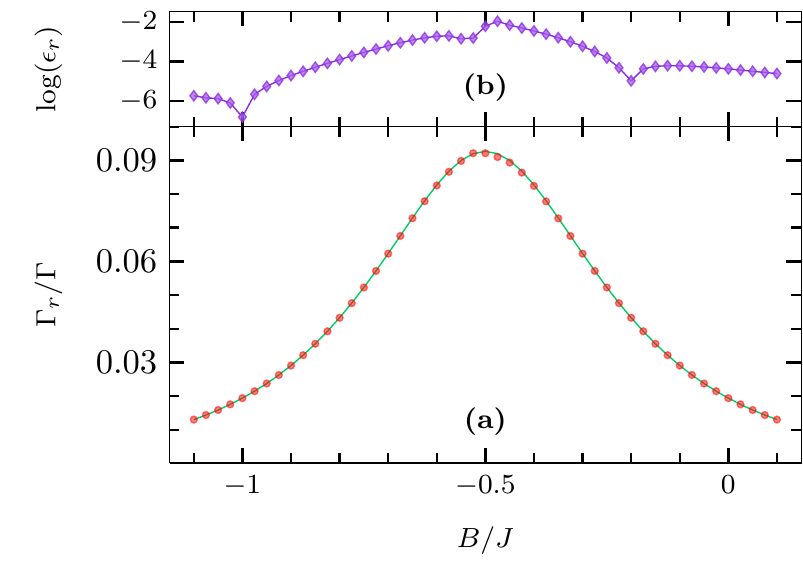}
}
\caption{(a) Comparison of the relaxation rates as a function of the energy
splitting $B$ determined either as the smallest non-zero real
eigenvalue $\Gamma_1$ of the Liouville superoperator (solid line), or as  
$\Gamma_r$ extracted from the Fourier spectrum at zero frequency (symbols). (b)
Relative difference $\epsilon_r=|\Gamma_r/\Gamma_1-1|$ between the two rates.
The parameters are $J=2\,\Gamma$, $\Delta=0.2\,\Gamma$, $eV=\Gamma$, and
$\beta=1.2/\Gamma$. }
\label{fig:rel}      
\end{figure}
%
%
In passing, we note that so far, we have determined the relaxation rate
$\Gamma_r$ as the smallest non-zero real eigenvalue according to
Eq.~\eqref{eq:denstev}. Alternatively, we may also consider the Fourier
transform of the spin relaxation dynamics and may extract the relaxation rate
as the zero-frequency value of the Fourier transform. Both ways yield
coinciding results as shown in Fig.~\ref{fig:rel} (a) as a function
of the energy bias $B$. The tiny numerical difference is shown in
Fig.~\ref{fig:rel} (b).

\subsection{Nonequilibrium Spin Dephasing}
\begin{figure}[t!]
\resizebox{0.49\textwidth}{!}{%
  \includegraphics{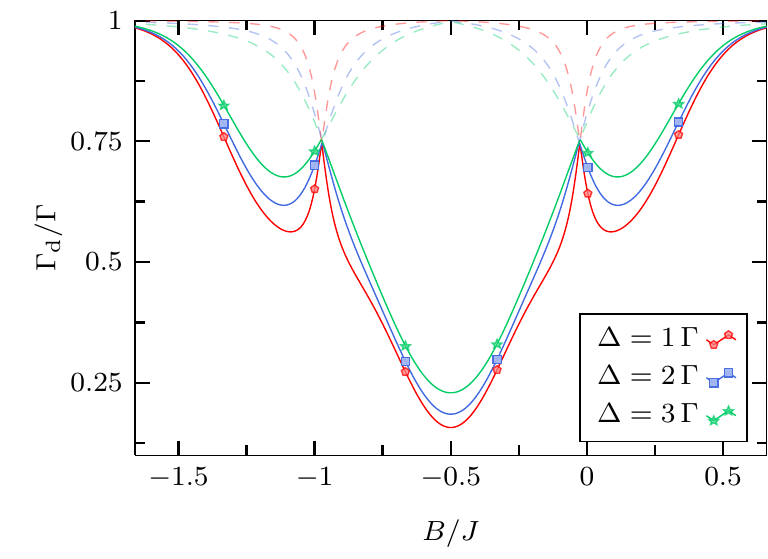}
}
\caption{Decoherence rate $\Gamma_d$ for different values of $\Delta$.
The decoherence rate is minimal at resonance when $B=-J/2$ and grows up to a
local maximum, when $B+J$ is one order of magnitude larger than $\Delta$.
The thick solid (blurred) lines show the real parts of the smallest (the second
smallest eigenvalue). The parameters are $J=15\,\Gamma$, $eV=10\,\Gamma$
and $\beta=0.7/\Gamma$.}
\label{fig:decB}      
\end{figure}

In addition to the pure relaxation dynamics at long times, the spin also shows
decaying oscillatory dynamics in Fig.\ \ref{fig:tev}. This spin dephasing 
 is generated by those eigenvalues of the Louvillian
superoperator $\bf W$, which have a non-zero imaginary part $\Omega_k\neq0$. 
The time scale on which the oscillations fade away is given by the
corresponding real part $\Gamma_k$. We consider in the following the
decoherence rate given by that eigenvalue with the smallest real part
$\Gamma_d$.  \\
The results for the decoherence rate as a function of the energy bias $B$ are
shown in Fig.~\ref{fig:decB} for different values of $\Delta$. Again, the
underlying spectrum of the system Hamiltonian is the basis to rationalize the
behavior of $\Gamma_d$. We find a constant decoherence rate in the
regions far away from any resonance, i.e., for $\lvert 2B+J\rvert\gg\Delta$, and
a global minimum at resonance when $B=-J/2$. At this resonance, two pairs of
almost degenerate energy states exists, see Fig.\ \ref{fig:avcr}, such that the
dynamics around this point is most robust against dephasing. The value of the
decoherence rate away from the crossing region is independent of $\Delta$, which
is clear since for $B,B+J\gg\Delta$, the off-diagonal elements of $H_S$ are
negligible in comparison to $B$, and yield only small corrections of the order
$\mathcal{O}(\Delta^2)$. Then, dephasing is solely determined by the
parametric coupling of the spin operator $S_z$ to the electron number operator
$n$ on the dot. In this limit, the total Hamiltonian is diagonal and the
nonequilibrium fluctuations only induce dephasing with a rate dominated by the
coupling $\Gamma$ to the reservoir fluctuations. Furthermore,
we observe exact degeneracies of the real parts of two different eigenvalues
around $B=-J$ and $B=0$ with additional corresponding minima in the
rate. Moreover, the shape is perfectly symmetric to
the point $B=-J/2$.
%
%
%
%

\begin{figure}[t!]
\resizebox{0.49\textwidth}{!}{%
  \includegraphics{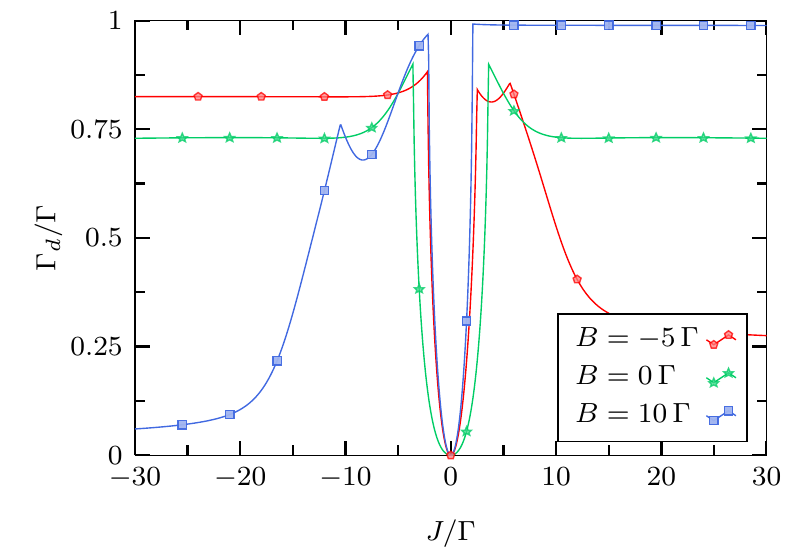}
}
\caption{Decoherence rate $\Gamma_d$ as a function of $J$ for different values
of the energy bias $B$ for the parameters 
$\Delta=3\,\Gamma$, $eV=10\,\Gamma$ and $\beta=0.7/\Gamma$.}
\label{fig:decJ}       
\end{figure}

Finally, we study the dependence of the decoherence rate on $J$. The results 
 for different values of the bias $B$ are shown in Fig.~\ref{fig:decJ}. For
$B=0$, the system is totally symmetric under an exchange of the states
$\ket{\uparrow}$ and $\ket{\downarrow}$, which is reflected in the
coherence behaviour as well. In all three
cases, the decoherence rate vanishes for $J=0$, since then, the spin is
decoupled from the electronic fluctuations. At first, for larger $|J|$, the
dephasing rate grows, until the smallest and the second smallest eigenvalues
cross. Beyond that degeneracy point, $\Gamma_d$ decreases again. The
decrease depends on the energy bias $B$ and on the relative sign of between~$J$
and~$B$. For $B=0$, $\Gamma_d$ is perfectly symmetric, while for $B\neq0$,
another local minimum around
$J\approx-B$ appears.
%
%

\subsection{Electron current and differential conductance}

\begin{figure}[t!]
\resizebox{0.49\textwidth}{!}{%
  \includegraphics{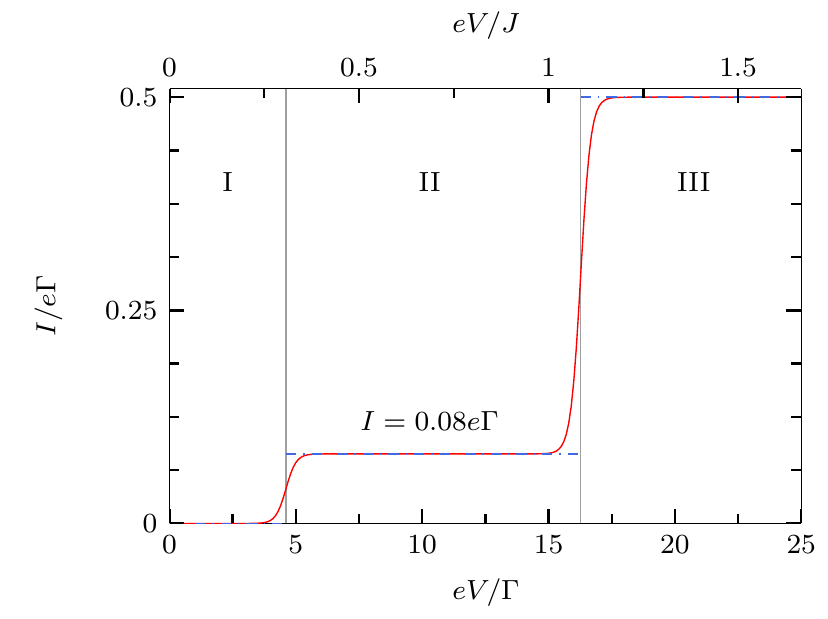}
}
\caption{Stationary current $I$ as a function of the bias voltage $eV$ for the
parameters $B=-5\,\Gamma$, $J=15\,\Gamma$, $\Delta=3\,\Gamma$ and
$\beta=10/\Gamma$. The vertical lines divide the plot into three regions, where
either no (I), one (II) or two (III) energy levels of the dot are in the energy
window spanned by the chemical potential. The dashed line is the value of the
current in region II given by Eq.\ \eqref{eq:currII}. }
\label{fig:current}       
\end{figure}

For completeness, we investigate the stationary current $I$ given
by Eq.~\eqref{eq:curr}. The current-voltage characteristics is shown in
Fig.~\ref{fig:current} and can clearly be divided into
three different regions. There is no current in the region I of very small
bias voltages. This extends until the voltage is large enough
such that the first transport channel opens. This happens when
$eV=2(\epsilon_0^--\epsilon_1^-)$, which is the energy needed to occupy the
quantum dot with one electron, while the spin is kept fixed. In the region II, 
the current reaches another plateau whose height depends on the dot parameters
$B$, $J$ and $\Delta$, simply because of the structure of the creation operator
$d$ in the eigenbasis of $ H_S$. An additional transport channel opens up when
the voltage is large enough to change the spin quantum number from $-$ to $+$.
This occurs at $eV=2(\epsilon_0^+-\epsilon_1^-)$. In the region III, the current
is given by $I^\text{III}=e\Gamma/2$, which is the maximum values for this
type of tunneling interaction. A finite temperature yields a smoothening of the
transition regions while for exactly zero temperature, perfect steps would
appear.\\
An estimate for the current in region II is found by using
equation~\eqref{eq:curr}. We calculate the trace of the matrix ${\bf W}^I$
assuming sharp steps in the Fermi function and weight everything with a factor
$1/4$. We find
\begin{equation}\label{eq:currII}
 I^\text{II}=\frac{e\Gamma}{4}\left[1+\frac{\Delta^2+B(B+J)}{\sqrt{
(\Delta^2+B^2)(\Delta^2+(B+J)^2)}}\right],
\end{equation}
which is in good agreement with the numerically calculated current in this
region.
%
%
%
%

\begin{figure}[t!]
\resizebox{0.49\textwidth}{!}{%
  \includegraphics{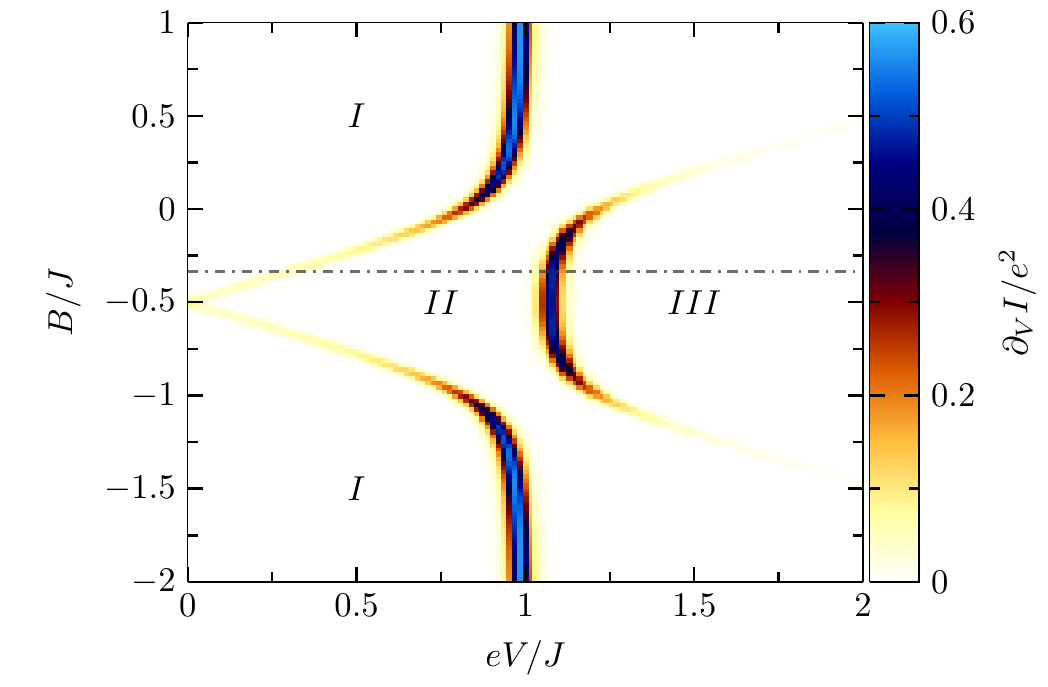}
}
\caption{Differential conductance $\partial I/\partial V$ as a function of $B$
and voltage $eV$ for the same parameters as in Fig.\ \ref{fig:current}. 
The dashed horizontal line indicates the value $B=-J/3$. }
\label{fig:conduct}       
\end{figure}

Finally, in addition to the current, we show the differential conductance
$\partial I/\partial V$ in Fig.~\ref{fig:conduct}. Near resonance
$B=-J/2$, we find two different lines on which the conductance is different from
zero. These lines correspond to $eV=2(\epsilon_0^--\epsilon_1^-)$ and
$eV=2(\epsilon_0^+-\epsilon_1^-)$, dividing the diagram into the three regions
I, II and III. Moving away from resonance, i.e., increasing or decreasing the
bias, the separatrix I-II becomes more pronounced, while the separatrix II-III 
fades out. This is clear since when we increase the voltage for a fixed $B$
around the resonance $B=-J/2$, the current first jumps to a small level and
only at even larger bias to a large value (an example for this is shown in
Fig.\ \ref{fig:current}). In turn, away from resonance ($B\gg 0$ or $B\ll-1$),
the current jumps directly almost to the maximal value when crossing the
separatrix I-II. Hence for still growing $V$, no further significant increase
is possible when crossing the separatrix II-III for a fixed value of $B$. The
reason is again rooted in the spectrum of the system Hamiltonian: the fine
structure of the spectrum in the crossing regions is only resolved when
$B\approx -J/2$. Outside of this region, the four avoided level crossings 
appear as one large ``global'' crossing. 
%
%
%
%

\section{Correlation Function of the Nonequilibrium Fluctuations}\label{sec:4}

When a physical system is subject to fluctuations which relax it to some
stationary or thermal equilibrium state, it is of foremost interest to know the
properties of the fluctuations in terms of their mean value and their higher
moments. They determine the relaxation (and dephasing) rates via their spectral
distribution. At thermal equilibrium and for a Gaussian environment, the
fluctuations are composed of a sum over harmonic thermal fluctuations with a
spectral weight given by the spectral density of the bath \cite{Weiss}.
Under nonequilibrium conditions, such simple features are not known up to
present. Here, we can calculate \cite{Chaudhuri99} the correlation properties of
the nonequilibrium fluctuations which act on the spin$-1/2$ and address the
questions how the well-known fluctuation-dissipation theorem of equilibrium
statistical physics is modified under nonequilibrium conditions. Moreover, it
is straightforward to generalize the results for a static bias voltage to
include also an ac voltage driving. 

\subsection{Static Bias Voltage}

We consider the Heisenberg equations of motion for spin-1/2 subsystem which are readily 
obtained as 
\begin{eqnarray}
\label{eq:spin1}
\dot{S}_x(t)&=&-B S_y(t) - J d^\dagger(t) d(t) S_y(t) \, , \\
\dot{S}_y(t)&=& B S_x(t) -\Delta S_z(t) + J d^\dagger(t) d(t) S_x(t)\, , \\
\dot{S}_z(t)&=&\frac{\Delta}{2} S_y(t) 
\end{eqnarray}
in the presence of the coupling to the electronic occupation of the dot. Due to the
tunneling coupling to the leads, the occupation number is a fluctuating
quantity whose correlation properties will be determined below. The dynamics of
the fluctuations is determined by the corresponding Heisenberg equations of
motion of the fermionic operators
\begin{eqnarray}
	\label{eq:ddot}
\dot{d}(t)&=&-i{J}S_{z}(t)d(t)-i\sum_{k\alpha}t_{k\alpha}c_{
 k\alpha}(t), \\
	\label{eq:cdot}\dot{c}_{k\alpha}(t)&=&-i\epsilon_{k\alpha}{c}_{k\alpha}
(t)-it_{k\alpha}d(t)\, .
\end{eqnarray}
They follow after neglecting terms that are at least cubic in
the creation and annihilation operators. This is consistent with the Markovian
approximation made in the preceeding sections. The latter implicitly only
includes sequential tunneling events between the reservoirs and the dot, i.e.,
 single-particle excitations in the system. 
We solve the inhomogeneous linear differential equation \eqref{eq:cdot} and find
\begin{equation*}
 c_{k\alpha}(t)=e^{-i\epsilon_{k\alpha}(t-t_{0})}\left[c_{k\alpha}(t)-it_{
k\alpha }
\int_{t_{0}}^t \!\!ds \,d(s)e^{i\epsilon_{k\alpha}(s-t_{0})}\right].
\end{equation*}
We insert this solution in Eq.\ \eqref{eq:ddot} and consider the time in
that $d(t)$ significantly changes to be rather large
compared to the free time evolution of the reservoir operators
$c_{k\alpha}(t)$. This is reasonable due to the continuous density of the
states of the reservoirs for which we may assume very rapid thermalization on
time scales very short as compared to any time scale of the system dynamics.
Again, this is consistent with the wide-band limit used throughout this work.
Then, we can evaluate the integral and arrive at
\begin{equation}
	\dot{d}(t)=-i{J}S_{z}(t)d(t)-\Gamma d(t)+{f}(t)\, ,\label{eq:ddot2}
\end{equation}
with 
\begin{equation}
 f(t)=-i\sum_{k\alpha}t_{k\alpha}e^{-i\epsilon_{k\alpha}(t-t_{0})}c_{k\alpha} 
\end{equation}
acting as a fluctuating force with vanishing mean value but a non-vanishing
correlation. Using once more the wide-band approximation with rates
$\Gamma_{\alpha}={2\pi}\sum_{k}|{t_{k\alpha}}|^2\delta(E-\epsilon_{k\alpha})$
gives rise to a prefactor $\Gamma=(\Gamma_L+\Gamma_R)/2$.\\
Equation \eqref{eq:ddot2} allows us now to determine the dynamics of
the electronic
degrees of freedom on the quantum dot under the influence of the nonequilibrium
fluctuating operator $f(t)$.
In the limit $t_0\rightarrow-\infty$, we find that the dynamics of the dot occupation become independent of the spin dynamics which is directly induced 
by the particle fluctuation.
Then, the occupation number is simply given by
\begin{equation}
 n(t)=\lim_{t_0\rightarrow-\infty}\int_{t_0}^td\tau_1d\tau_2f^\dagger(\tau_1)f(\tau_2)e^{\Gamma(\tau_1+\tau_2-2t)},
\end{equation}
with $n(t)=d^\dagger(t)d(t)$.

\subsubsection{Nonequilibrium noise characteristics}
Equipped with this, we are able to calculate the characteristics of the nonequilibrium quantum noise. In the limit of symmetric tunneling, the mean value follows readily as
\begin{equation}
\langle n(t)\rangle\equiv \gamma^+(0)=\frac{1}{2}\,.
\label{eq:noisemean}
\end{equation}
The autocorrelation function of the dot occupation can be calculated as
\begin{equation}
L(t,s)= \langle n(t)n(s)\rangle- \langle n(t)\rangle\langle n(s)\rangle=\gamma^+(t-s)^2
\label{eq:correlation}
\end{equation}
Moreover, we have
defined the functions
\begin{equation} \gamma_\alpha^\pm(t)=\frac{\Gamma_\alpha}{2\pi}\int
{d}E \frac{f^\pm_\alpha(E)}{E^2+\Gamma^2}e^{iEt}\, ,
\end{equation}
and $\gamma^\pm(t)=\sum_\alpha \gamma_\alpha^\pm(t)$. Here, 
$f^\pm_\alpha(E)=[1+e^{\pm\beta(E-\mu_\alpha)}]^{-1}$ is the Fermi distribution.
The correlation function is translation invariant in time and we can simplify our notation 
according to $L(t,s)=L(t-s,0)\equiv L(t-s)$.

The spectral resolution of the nonequilibrium 
quantum noise can be obtained via the Fourier transform $L(\omega)=\int dt L(t)e^{-i\omega t}$. Simple expressions follow in the limit of zero and infinite temperature. 

In the case of infinite temperature, we find 
\begin{equation}
  \lim_{\beta\rightarrow0}L(\omega)=\frac{1}{2}\frac{2\Gamma}{\omega^2+4\Gamma^2} \,,
\end{equation}
while the limit of zero temperature amounts to 
%


\begin{align}
\begin{split}
\lim_{\beta\rightarrow\infty}L(\omega)=&\frac{1}{2}\frac{2\Gamma}{\omega^2+4\Gamma^2}\biggl[\mathcal{C}(\omega,eV)+\mathcal{C}(\omega,-eV)\biggr]\,,
\end{split}
\end{align}

with
\begin{widetext}
\begin{equation}
\mathcal{C}(\omega,eV)=\frac{1}{\pi}\left[\Theta(\omega)+\Theta(\omega-eV)\right]\biggl[\atan\biggl(\frac{2\omega-eV}{2\Gamma}\biggr)-\atan\biggl(\frac{eV}{2\Gamma}\biggr)+\frac{\Gamma}{\omega}\log\biggl(\frac{\Gamma^2+(\omega+eV/2)^2}{\Gamma^2+(eV/2)^2}\biggr)\biggr] \,.
\end{equation}
\end{widetext}

\begin{figure}[t!]
\resizebox{0.5\textwidth}{!}{%
  \includegraphics{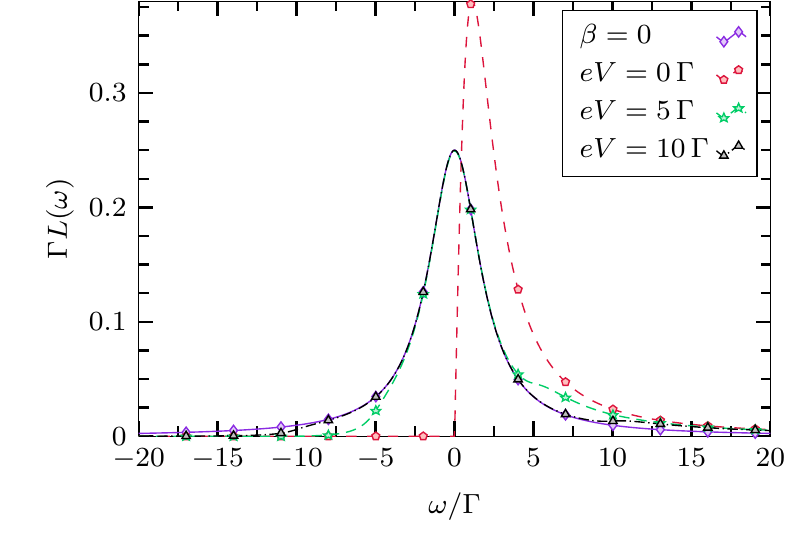}
}
\caption{Correlation function $L(\omega)$ for infinite temperature (diamond) and zero temperature for three different voltages (pentagon, star, triangle).}
\label{fig:corr}       
\end{figure}

For compactness, we present only the result for $\Gamma=\Gamma_L=\Gamma_R$ and a symmetric voltage between both leads. However, the result for $T\to 0$ within the sequential tunneling limit ($T>\Gamma$) is general with $\Gamma=(\Gamma_L+\Gamma_R)/2$. 
Further, we study the limit of a large voltage $eV\gg\Gamma$ and find that
in this limit the infinite and zero temperature solution are identical, i.e., $\lim_{eV\rightarrow\infty}\lim_{\beta\rightarrow\infty}L(\omega)=\lim_{\beta\rightarrow0}L(\omega)$.
In figure~\ref{fig:corr}, we show $L(\omega)$ for infinite (solid line) and zero (dashed lines)  temperature for different voltages. In all cases, the spectral resolution of the nonequilibrium noise correlation function is dominated by a Loren\-tzian centered at zero frequency. In the zero temperature limit, a shoulder appears 
close to $\omega=eV$, whereas on the opposite position below $\omega=-eV$ the spectral weight is reduced.

In general, the exact expression at arbitrary temperature involves sums over Matsubara frequencies, which cannot be evaluated in a closed expression. Yet, these sums generate peaks in the spectral noise function at $\omega=\pm eV/2$ and $\omega=\pm eV$ with temperature dependent widths determined by the Matsubara frequencies.
%

%
%
%
%
\subsubsection{Nonequilibrium fluctuation relation}
According to the fluctuation dissipation theorem (FDT) valid at quantum statistical equilibrium,  an equilibrium correlation
function $L_{\rm eq}(t)$ satisfies the relation $L_{\rm eq}(t-i\beta)=L_{\rm eq}^*(t)$. Checking this
for the nonequilibrium occupation correlation to study the difference in the
structure, we find
\begin{equation}
\gamma^+_\alpha(t-i\beta)=e^{\beta\mu_\alpha}\gamma^-_\alpha(t)
\end{equation}
and
\begin{equation}
 \gamma^+_L(t)=[\gamma^-_R(t)]^*.
\end{equation}


Using these two identities, we can calculate the final expression
\begin{align}
L(t-i\beta)=L^*(t)-\sum_\alpha\left(1-e^{-2\beta\mu_\alpha}\right)[\gamma_{\alpha}^+(t)]^*[\gamma_{\alpha}^+(t)]^*\, ,
\label{eq:fdtnd}\end{align}
which extends the equilibrium FDT to the nonequilibrium regime. Note that in the equilibrium case of vanishing voltage $\mu_\alpha=0$, the equilibrium FDT is recovered.

\subsection{Time-dependent AC-Voltage}\label{sec:5}
Similar results follow for ac-driven leads. We exchange the energies of the lead
fermions by $\epsilon_{k\alpha}(t)=\epsilon_{k\alpha}+eV_\alpha\cos{(\Omega
t)}$. Using a time-dependent unitary transformation
$U(t)=e^{-i\sum_\alpha\phi_\alpha(t)N_\alpha}$ with
$\phi_\alpha(t)=eV_\alpha\int_{t_0}^t d\tau \cos{(\Omega \tau)}$ we transfer the
time dependency from the energy $\epsilon_{k\alpha}(t)$ to the tunnel amplitude
$t_{k\alpha}(t)=t_{k\alpha}e^{i\phi_\alpha(t)}$. Here, $N_\alpha$ is the particle number operator for the lead labelled by $\alpha$.
Hence, by making use of the 
Jacobi-Anger identity to expand the exponential
\begin{equation}
e^{iz\sin{\theta}}=\sum_{n}J_n(z)e^{in\theta}
\end{equation}
in terms of Bessel-functions $J_n(z)$, we follow the same calculation as before.
We find for the ac-driven correlation function 
\begin{align}
\tilde{L}(t,s)&=\sum_{kl}\sum_{mn}\gamma^+_{kl}(t-s)\gamma^+_{mn}(t-s)e^{i(k+m)\Omega t}e^{-i(l+n)\Omega s},
\label{eq:corrd}\end{align}
where we have defined in analogy to the non-driven case
\begin{align}\begin{split}
2\pi{\gamma}^\pm_{k,l,\alpha}(t)&=\Gamma_\alpha
J_k(eV_\alpha/\Omega)J_{l}(eV_\alpha/\Omega)\\
&\!\!\times\!\!\int{d}E\;
\frac{f_\alpha^\pm(E)e^{iEt}}{\Gamma^2+(E+k\Omega)(E+l\Omega)+i(k+l)\Omega\Gamma}
\end{split}\end{align}
and again $\sum_\alpha \gamma_{m,n,\alpha}^\pm(t)=\gamma_{m,n}^\pm(t)$.
By averaging over one period $T=2\pi\Omega^{-1}$, we regain the translational invariance
\begin{equation}
 L(t-s)=\frac{1}{T}\int_0^T d\tau \tilde{L}(t+\tau,s+\tau),
\end{equation}
which reduces the correlation function to
\begin{align}
 L(t)&=\sum_{klm}\gamma_{k-m,l}^+(t)\gamma_{m,k-l}^+(t)e^{ik\Omega t}.
\end{align}

Now, to check the fluctuation dissipation relation, we make use of the following identities
\begin{equation}
{\gamma}_{k,l,\alpha}^+(t-i\beta)=e^{\beta(\mu_\alpha-m\Omega)}{\gamma}_{k,l,\alpha}^-(t)
\end{equation}
and
\begin{equation}
 \sum_{kl}\gamma_{k,l,R}^+(t)=\sum_{kl}[\gamma_{k,l,L}^-(t)]^*.
\end{equation}
Finally, we arrive at
\begin{align}\begin{split}
 L(t-i\beta)&=L^*(t)-\sum_\alpha\sum_{klm}(1-e^{-2\beta\mu_\alpha})\\&\times
[\gamma^+_{k-m,l,\alpha}(t)]^*[\gamma_{m,k-l,\alpha}^+(t)]^*e^{-ik\Omega t}.
\end{split}\end{align}

\section{Conclusions}
%
By taking the point of view of quantum transport through a single-level quantum dot which is coupled to a quantum two-state system, we have analyzed the real-time dynamics of the two-state system under the action of nonequilibrium quantum statistical fluctuations. Under the assumption of weak dot-lead electronic tunneling, a diagrammatic perturbation method on the basis of sequential charge tunneling can be used. The coupling between the electronic level and the two-state system can be arbitrary. A finite voltage between the electron reservoirs easily allows to generate nonequilibrium fluctuations. The electronic degrees of freedom can be integrated over such that the quantum dissipative dynamics of the two-state system itself can be easily studied. We have concentrated on the relaxation and dephasing processes at long time by calculating the smallest nonzero eigenvalues of the underlying Liouville operator. In the regime of sequential tunneling processes, the action of the quantum noise is sufficiently weak such that the unperturbed energy spectrum of the dot-plus-two-state-system is a useful starting point. Since the eigenvalues of a Liouvillian rate matrix can easily be evaluated numerically, we have straightforward access to the relaxation and dephasing rates as a function of the various model parameters. We find a rich structure in both these observables which can consistently being traced back to the unperturbed energy spectrum. 

Furthermore, an analysis in terms of Heisenberg-Lan\-ge\-vin equations of motion allows us to extract the autocorrelation function of the nonequilibrium quantum statistical noise in the limit of a Markovian approximation. The zero-temperature limit allows us to obtain a simple spectral decomposition of the frequency components of the nonequilibrium noise under the action of a static dc voltage. Likewise, a generalized nonequilibrium fluctuation relation follows, which reproduces the well-known equilibrium fluctuation-dissipation theorem at zero voltage. A generalization to time-periodic ac voltages is also possible, such that a generalized driven fluctuation relation results which involves all higher harmonics.

\begin{acknowledgement}
We thank P.\ Nalbach for fruitful discussions. Moreover, we acknowledge support by the DFG SFB 668 ``Magnetismus vom Einzelatom zur Nanostruktur''. 
\end{acknowledgement}

\
\end{document}